\documentclass[%
 aip,
 cha,
 amsmath,amssymb,
 floatfix,
 reprint,%
]{revtex4-1}
\usepackage[utf8]{inputenc}
\usepackage{graphicx}
\usepackage{dcolumn}
\usepackage{bm}

\usepackage[T1]{fontenc}
\usepackage{mathptmx}
\usepackage{etoolbox}
\usepackage{booktabs}
\makeatletter
\def\@email#1#2{%
 \endgroup
 \patchcmd{\titleblock@produce}
  {\frontmatter@RRAPformat}
  {\frontmatter@RRAPformat{\produce@RRAP{#1\href{mailto:#2}{#2}}}\frontmatter@RRAPformat}
  {}{}
}%
\makeatother
\makeatletter
\def\@bibdataout@aip{%
 \immediate\write\@bibdataout{%
  @CONTROL{%
   aip41Control%
   \longbibliography@sw{\true@sw}{\aip@jtitx@sw{\false@sw}{\true@sw}}%
   {%
    ,author="48",pages="1",title="0"%
   }{%
    ,author="48",pages="0",title=""%
   }%
  }%
 }%
 \if@filesw
  \immediate\write\@auxout{\string\citation{aip41Control}}%
 \fi
}%
\makeatother
\newcommand{\manualfigurecaption}[2]{\refstepcounter{figure}\label{#1}\par\smallskip\noindent\parbox{\linewidth}{\small\textbf{FIG. \thefigure.} #2}\par\normalsize}
\newcommand{\manualwidefigurecaption}[2]{\refstepcounter{figure}\label{#1}\par\smallskip\noindent\parbox{\textwidth}{\small\textbf{FIG. \thefigure.} #2}\par\normalsize}
\newcommand{\manualtablecaption}[2]{\refstepcounter{table}\label{#1}\par\smallskip\noindent\parbox{\linewidth}{\small\textbf{TABLE \thetable.} #2}\par\smallskip\normalsize}
\newcommand{\manualwidetablecaption}[2]{\refstepcounter{table}\label{#1}\par\smallskip\noindent\parbox{\textwidth}{\small\textbf{TABLE \thetable.} #2}\par\smallskip\normalsize}
\begin{document}

\preprint{AIP/123-QED}

\title[OCVF for Many-Body System]{Infusing Experimental Reality into Complex Many-Body Hamiltonians: The Observable-Constrained Variational Framework (OCVF)}
\author{ZP. Yang}
\altaffiliation{These authors contributed equally to this work.}
\affiliation{National Key Laboratory of Human-Machine Hybrid Augmented Intelligence, National Engineering Research Center for Visual Information and Applications, and Institute of Artificial Intelligence and Robotics, Xi'an Jiaotong University, Xi'an, 710049, China.}
\author{SL. Guo}
\altaffiliation{These authors contributed equally to this work.}
\affiliation{
School of Microelectronics\&StateKey
 Laboratory forMechanical Behavior of Materials,Xi'an
 JiaotongUniversity,Xi'an710049,China;KeyLaboratoryof
 Micro-NanoElectronicsand System Integration of Xi'an City,
 Xi'anJiaotongUniversity,Xi'an710049,China
}
\author{Nan Zhang}
\email{nzhang1@xjtu.edu.cn}
\altaffiliation{Corresponding author.}
\affiliation{National Key Laboratory of Human-Machine Hybrid Augmented Intelligence, National Engineering Research Center for Visual Information and Applications, and Institute of Artificial Intelligence and Robotics, Xi'an Jiaotong University, Xi'an, 710049, China.}
\affiliation{Electronic Materials Research Laboratory, Key Laboratory of the Ministry of Education \& International Center for Dielectric Research, School of Electronic Science and Engineering, Xi'an Jiaotong University, Xi'an, 710049, China.}
\author{Ping Wei}
\affiliation{National Key Laboratory of Human-Machine Hybrid Augmented Intelligence, National Engineering Research Center for Visual Information and Applications, and Institute of Artificial Intelligence and Robotics, Xi'an Jiaotong University, Xi'an, 710049, China.}
\author{Nanning Zheng}
\affiliation{National Key Laboratory of Human-Machine Hybrid Augmented Intelligence, National Engineering Research Center for Visual Information and Applications, and Institute of Artificial Intelligence and Robotics, Xi'an Jiaotong University, Xi'an, 710049, China.}

\date{\today}

\begin{abstract}
Machine learning potentials have revolutionized simulations of complex many-body systems, yet their predictive accuracy is fundamentally limited by the quality of the Density Functional Theory (DFT) data used for training. Systematic DFT errors, especially in entropy and long-range interaction effects, can lead to incorrect predictions of macroscopic phase transitions. To address this bottom-up limitation, we propose the Observable-Constrained Variational Framework (OCVF), a top-down correction paradigm. In OCVF, the DFT-trained potential is retained as a transferable structural skeleton, while a neural correction term (the physical flesh) is learned from macroscopic experimental observables. By formulating molecular dynamics as a fully differentiable process and constraining the simulated and experimental pair distribution functions (PDFs), we optimize the effective Hamiltonian toward experimental reality. Using BaTiO$_3$ as a benchmark, OCVF corrects erroneous transition temperatures from the prior DFT-based model, improves the Cubic--Tetragonal transition accuracy by 95.8\%, improves Orthorhombic--Rhombohedral transition accuracy by 36.1\%, and recovers the correct C--T--O--R phase sequence with improved low-temperature robustness. These results establish a rigorous pathway for calibrating first-principles-derived models against experimental observations without sacrificing their generalization capability.
\end{abstract}

\maketitle

\section{\label{sec:introduction}Introduction:\protect\\The Origen of the problem}

Predicting macroscopic responses of complex many-body materials, especially ferroelectric phase transitions, remains a central challenge in computational physics: in BaTiO$_3$ (BTO), the characteristic Cubic--Tetragonal--Orthorhombic--Rhombohedral (C-T-O-R) sequence arises from a delicate coupling among interatomic interactions, anharmonicity, and entropy \cite{Ouyang2023,zhong1994phase,zhong1995first}.

Traditional strategies span classical force fields that require substantial physical intuition and laborious parameter fitting \cite{zeng2011thermo,wu2017electric,sepliarsky2004atomistic,asthagiri2006advances,padilla1996first}, effective-Hamiltonian-based first-principles frameworks and their modern data-driven extensions \cite{zhong1994phase,zhong1995first,ma2025active}, and ab initio molecular dynamics (AIMD), whose computational cost limits simulations to relatively small systems \cite{Ouyang2023,liu2018understanding,wang2016subterahertz,wang2014fano,wang2012atomistic,wang2011fermi}. Since phase-transition calculations typically require much larger cells and longer timescales and complex phase transition mechanism\cite{zhou2022strain}, an evident scale gap remains; to bridge this gap, bottom-up machine learning potentials (MLPs) based on neural-network descriptors have become a practical route, as demonstrated in recent BTO studies by Zhang et al. \cite{Ouyang2023,zhang2022structural}. The core of this bottom-up paradigm is to train a neural network to reproduce the high-precision potential energy surface (PES) generated from Density Functional Theory (DFT) calculations . Outstanding recent work, such as that by Ouyang et al. \cite{Ouyang2023}, has demonstrated the power of this approach. Using an MPNN (a modified DimeNet++ \cite{gasteiger2003directional}), they successfully constructed a high-precision BTO potential by fitting to the energies, forces, and stresses calculated by DFT (PBE functional). Their model, under an NPT ensemble, successfully reproduced the C-T-O-R phase transition sequence of BTO \cite{Ouyang2023}. However, a problem remains: systematic deviation in the "skeleton" model persists. This bottom-up success exposes an unavoidable, fundamental problem: the accuracy of the MLP is shackled by the accuracy of its training data (i.e., DFT) \cite{Ouyang2023}. The neural network's accuracy is capped by its training set; it faithfully learns everything from the DFT data, including DFT's own inherent systematic biases . The work of Ouyang et al. clearly and even quantitatively, demonstrates this: they found that the BTO phase transition temperatures are predicted from their "bottom-up" training were offset,the simulation of polarization phase transition obtained from training with DFT exhibits severe distortion under the NVT ensemble, as shown in FIG. \ref{fig:results of DFT} [a]. However, under the NPT ensemble, the polarization phase transition remains consistent with the lattice phase transition, as illustrated in FIG. \ref{fig:results of DFT} [b], [c], and [d],with C-T at 475K, T-O at 275K, and O-R at 100K \cite{Ouyang2023}. However, when we applied the same method mentioned above to perform simulations in an NPT ensemble environment with parameters set as timestep = 0.05 * units.fs, ttime = 50 * units.fs, and pfactor = 1e6 * units.GPa * (units.fs ** 2), we failed to obtain the approximate results. Moreover, catastrophic non-physical results emerged at low temperatures, which demonstrates that the initial model trained solely by DFT lacks robustness,as shown in FIG. \ref{fig:results of DFT}[e],[f].

We address this limitation by proposing the Observable-Constrained Variational Framework (OCVF), a top-down correction strategy that complements, rather than replaces bottom-up learning. Unlike standard bottom-up approaches that primarily fit microscopic energies and forces, OCVF follows a physical "skeleton-and-flesh" formulation: (i) the DFT-trained MLP is retained as a fundamental structural backbone $H_o$, preserving essential chemical topology and short-range repulsion that are difficult to infer from sparse experimental data; and (ii) a neural correction term $\Delta H_{\theta}$ is introduced into the fundamental prior bottom-up work and trained by macroscopic experimental observables $\mathfrak{O}_{\text{exp},s}$, thereby capturing missing physics (e.g., long-range correlations and entropy effects) not fully described by DFT \cite{Ouyang2023}.

Although conceptually simple, constructing and training $\Delta H_{\theta}$ is nontrivial and facing with many technical problems(e.g., Cuda out of Memory). Thus, in this work, we combine differentiable molecular dynamics with Neural Ordinary Differential Equations (NODEs) to enable this optimization and effectively invert the statistical-mechanical mapping: the microscopic Hamiltonian is iteratively adjusted until simulated macroscopic responses align with experimental reality \cite{chen2020learning}.
\refstepcounter{figure}\label{fig:results of DFT}
\begin{widetext}
\begin{center}
\includegraphics[width=\linewidth]{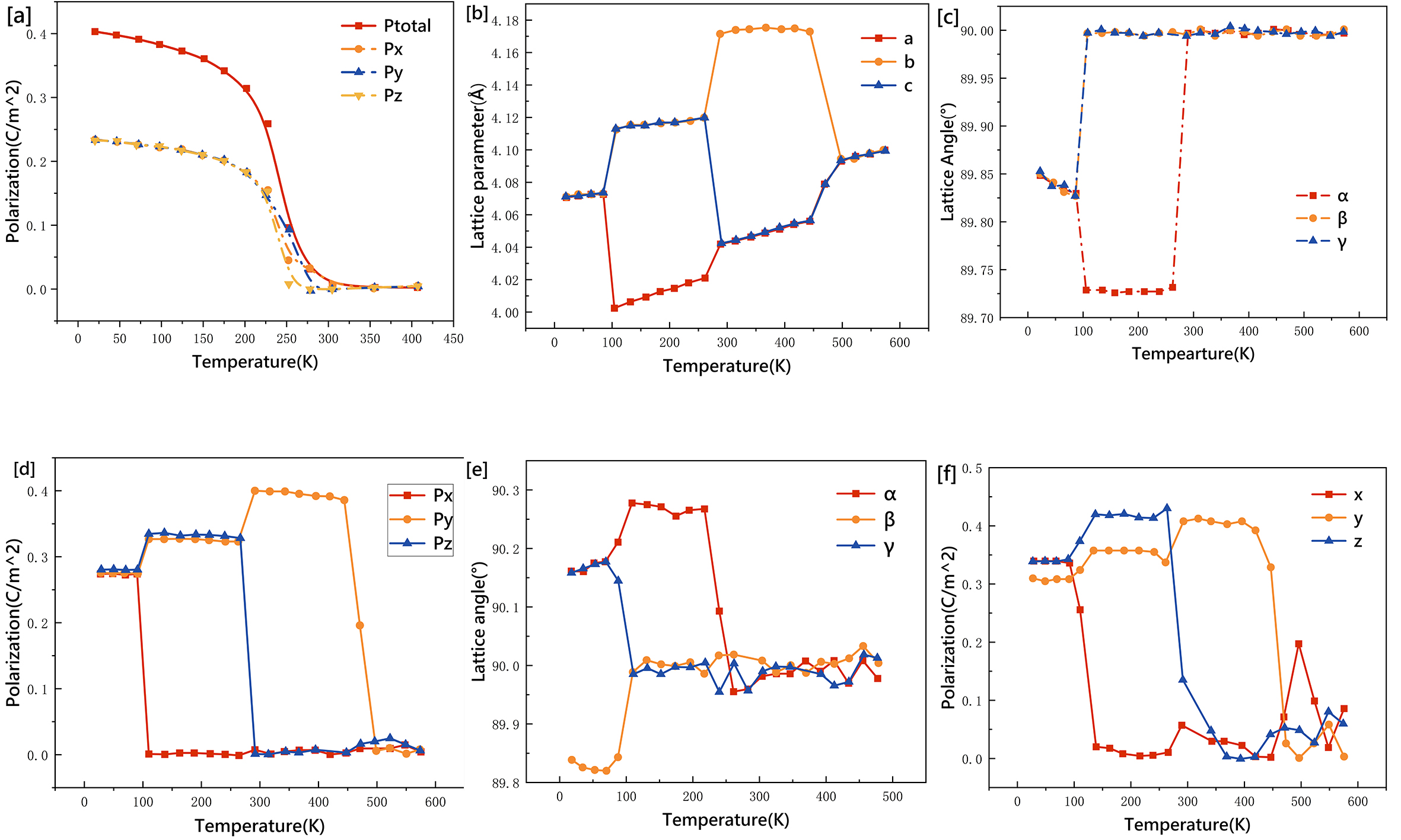}\par\smallskip
\small\textbf{FIG. \thefigure.} The results of training machine learning potential energy surfaces based on DFT.
\end{center}
\end{widetext}
\section{\label{sec:method}METHOD}

Basing on statistical physics and the specific context of this problem, this solving framework rests on seven fundamental assumptions:
\subsection{\label{subsec:core_stat_mech_assumptions}Core Statistical-Mechanical Assumptions}   
 Both the real system and the prior system are in thermal equilibrium, with the probability distribution of their microscopic states following the Boltzmann form (Gibbs distribution): $\rho \propto \exp(-\beta H)$ \cite{chandler1987introduction}. This is a core assumption of classical statistical mechanics, implying that the system’s macroscopic properties do not evolve over time and that the distribution of microscopic states depends solely on energy and temperature. Equivalence of Ensemble Averages and Macroscopic Observables: Ensemble averages are equivalent to macroscopic observables. The macroscopic observable $\mathfrak{O}_{\text{exp},s}$ is strictly equal to the ensemble average of its microscopic operator $\hat{O}_s$ in the real ensemble \cite{chandler1987introduction}—for example, the radial distribution function (PDF), which represents the statistical result of microscopic particle distances.Assumption of System Description: The state of the system is completely described by classical phase-space coordinates ($\mathbf{q}$, $\mathbf{p}$) and the Hamiltonian $H(\mathbf{q}, \mathbf{p})$ \cite{chandler1987introduction}. The Hamiltonian encodes all information about the system’s interactions (e.g., bonded and non-bonded interactions) and is sufficiently well-defined to specify the ensemble distribution: $\rho \propto \exp(-\beta H)$.

\subsection{\label{subsec:universal_function_approximator}Universal Function Approximator Assumptions }
  Universal Function Approximator: The neural network correction term $\Delta H_{\theta}$ (associated with $\Delta H$) can approximate complex functionals with arbitrary precision. This is based on the universal approximation theorem \cite{cybenko1989approximation}: feedforward neural networks especially Graph Neural Network with sufficiently many hidden layers can approximate any continuous function, enabling them to fit non-local, high-dimensional correction terms\cite{chen2019graph}.
\subsection{\label{subsec:credible_approximation}Credible Approximation Assumptions}
 The prior Hamiltonian $E_{\text{prior}}$ (e.g., DFT-PBE) is a reasonable approximation of the true Hamiltonian $\mathfrak{H}$. Specifically, the difference between the prior distribution $\rho_{prior}$ and the true distribution $\rho_{true}$ is a “local perturbation” rather than a “fundamental error” \cite{Ouyang2023}. If the prior deviation is excessive (e.g., using simple potential functions to simulate complex systems), the correction term may fail to effectively compensate for systematic errors. Rationality of Minimum Relative Entropy: Using the relative entropy $D_{\text{KL}}(\rho || \rho_o)$ to measure the difference between the corrected distribution $\rho_c$ and the prior distribution $\rho_o$ provides a strict definition of “minimal deviation. ”This assumption is rooted in information theory \cite{shell2008relative}: KL divergence is the minimal measure of information loss when approximating $\rho_c$ with $\rho_o$, ensuring that the corrected distribution retains the maximum information from the prior. Separability of Experimental Constraints: The microscopic operator $\hat{O}_s$ corresponding to the experimental observable $\mathfrak{O}_{\text{exp},s}$ is separable. In other words, the correction term $\Delta H = -k_B T \sum_s \lambda_s \hat{O}_s$ can be expressed as a linear superposition of individual operators. This assumption naturally arises from the Lagrange multiplier method, requiring that operators corresponding to experimental constraints are either independent or can be covered by linear combinations.

\section{\label{sec:bridging_constraints_hamiltonians}Bridging Experimental Constraints and Machine-Learning Hamiltonians}

Our goal is not to replace the prior Hamiltonian $H_0$ from scratch, but to calibrate it using experimental constraints so that the simulated ensemble reproduces the correct temperature-dependent structural statistics and phase-transition sequence. For BaTiO$_3$ (BTO), this distinction is essential: the prior model already contains the basic instability topology required for phase transitions, yet it does not yield quantitatively correct transition temperatures or phase boundaries over the full temperature range. This indicates that the main deficiency lies not in the absence of a transition mechanism, but in the inaccurate free-energy balance among competing phases.

The central idea of our framework is therefore to bridge experiment and machine learning at the ensemble level. Instead of directly fitting a neural network to another theoretical energy surface, we constrain the corrected model by experimental observables. Let $\rho_{\mathrm{exp}}(\mathbf{q},\mathbf{p})$ denote the true equilibrium ensemble under experimental conditions, and let $\mathfrak{O}_{\mathrm{exp},s}$ denote a macroscopic observable measured at thermodynamic state $s=(T,P,\mu,E_{\mathrm{field}},\dots)$. In this work, the primary observable is the pair distribution function (PDF), which serves as a structural correction of the underlying ensemble. The key requirement is that the corrected model should generate a simulated ensemble whose PDF matches the experimental PDF under the same NPT conditions.

From this perspective, we seek a corrected ensemble $\rho_c$ that remains as close as possible to a credible prior ensemble $\rho_0$ while satisfying experimental constraints \cite{jaynes1957information,shell2008relative}. In the classical constrained-ensemble formulation, the correction is expressed as a linear combination of microscopic operators associated with the chosen observables. Although this result provides an important physical foundation, it is too restrictive for complex many-body systems, where the discrepancy between the prior model and reality generally cannot be represented by a small set of linear observable operators(e.g.,$\hat\delta(r - |\mathbf{R}i - \mathbf{R}j|)$).

We therefore generalize this idea into the Observable-Constrained Variational Framework (OCVF). Instead of solving for a low-dimensional set of Lagrange multipliers, we introduce a neural correction $\Delta H_\theta$ and determine its parameters by minimizing the mismatch between simulated and experimental observables:
\begin{equation}
\begin{split}
L(\theta)&=\sum_{s=1}^{S} D_s\!\left(\mathfrak{O}_{\mathrm{sim},s},\mathfrak{O}_{\mathrm{exp},s}\right)\\
&=\sum_{s=1}^{S} D_s\!\left(F_s[H_0+\Delta H_\theta,\mathbf{Z}_s],\mathfrak{O}_{\mathrm{exp},s}\right)
\end{split}
\end{equation}
where $F_s$ denotes the differentiable forward model under thermodynamic condition $\mathbf{Z}_s$, and $D_s$ measures the discrepancy between simulated and experimental observables. In our implementation, $F_s$ consists of a differentiable NPT molecular dynamics simulation driven by the corrected Hamiltonian, followed by a differentiable PDF observer that maps trajectories to structural observables.

A crucial point is that the quantity learned here should not be interpreted as a bare Hamiltonian that explicitly changes with temperature. Instead, what we construct is a \emph{temperature-dependent effective Hamiltonian family}:
\begin{equation}
H_{\mathrm{eff}}(\mathbf{q},\mathbf{p},\mathbf{h};T)
=
H_0(\mathbf{q},\mathbf{p},\mathbf{h})
+
\Delta H_{\mathrm{eff}}(\mathbf{q},\mathbf{p},\mathbf{h};T),
\end{equation}
where $\mathbf{h}$ denotes the cell degrees of freedom in the NPT ensemble. The temperature dependence here does not mean that the underlying microscopic law itself is assumed to vary with $T$. Rather, it reflects the fact that different temperature regimes require different \emph{effective} corrections in order to reproduce the experimentally constrained ensemble statistics and the associated free-energy competition among phases.

In practice, we train correction networks at several temperature anchors, $T_k \in \{150,200,250,300\}\,\mathrm{K}$, and combine them using Gaussian temperature weights, enabling interpolation not only at the anchor points $T_k$ but also at intermediate thermodynamic states:
\begin{equation}
\Delta H_{\mathrm{eff}}(\mathbf{q},\mathbf{p},\mathbf{h};T)
=
\sum_{k=1}^{K} w_k(T)\,\Delta H_{\theta_k}(\mathbf{q},\mathbf{p},\mathbf{h}),
\end{equation}
with
\begin{equation}
w_k(T)=
\frac{\exp\!\left[-(T-T_k)^2/(2\sigma^2)\right]}
{\sum_j \exp\!\left[-(T-T_j)^2/(2\sigma^2)\right]}.
\end{equation}
This construction localizes each correction to its relevant temperature regime and suppresses destructive overlap between corrections learned from different thermodynamic conditions, while $\sigma$ ensuring a smooth interpolation over the full temperature range.

Importantly, each single temperature-anchored model $H_0+\Delta H_{\theta_k}$ is already capable of producing phase-transition behavior, which shows that the prior Hamiltonian retains the basic transition skeleton of BTO. However, no single anchored correction yields quantitatively accurate phase boundaries across the full range from 20 to 550 K. This observation motivates the Gaussian-stitched effective Hamiltonian family: the role of OCVF is not to create phase transitions artificially, but to recalibrate the temperature-dependent free-energy competition among phases through experimentally constrained ensemble correction.

With this construction, the gradient of the loss with respect to the neural correction parameters is obtained through the chain rule,
\begin{equation}
\frac{\partial L}{\partial \theta}
=
\sum_{s=1}^{S}
\frac{\partial D_s}{\partial \mathfrak{O}_{\mathrm{sim},s}}
\cdot
\frac{\partial F_s}{\partial H_{\mathrm{eff}}}
\cdot
\frac{\partial \Delta H_\theta}{\partial \theta},
\end{equation}
where the first term is the observational gradient, the second term is the physical gradient propagated through the differentiable NPT dynamics, and the third term is the model gradient obtained by automatic differentiation of the neural network. In this way, OCVF provides the bridge from classical ensemble correction to a practical machine-learning framework constrained directly by experimental PDF data.
\section{\label{sec:apply_ocvf_specific_model}Use OCVF to correct a specific model}

Through the analysis above, we arrive at an important conclusion: based on ensemble correction methods in quantum statistical mechanics, a corrected Hamiltonian $H_c$ derived from macroscopic experimental observables must exist \cite{jaynes1957information}. Its form is precisely that of the prior Hamiltonian $H_o$ superimposed with a correction field composed of a linear superposition of operators $\hat{O}_s$.

However, directly solving for this correction term is computationally prohibitive and analytical solutions are nearly impossible. Instead, the deviation in the Hamiltonian can be corrected by imposing the constraint that the macroscopic mean of the linear operator $\hat{O}_{sim}$ satisfies $\langle \hat{O}_{sim} \rangle_{\rho_{c}} = \mathfrak{O}_{exp, s}$, where $\langle \hat{O}_{sim} \rangle_{\rho_{c}}$ is the simulated observable and $\mathfrak{O}_{exp, s}$ is the true experimental observable. This is a foundational concept in inverse statistical mechanics, where macroscopic observables are used to infer microscopic potentials. A deviation in the Hamiltonian will be reflected in the moments of the phase-space distribution of $q$ and $p$. Therefore, correcting the $q/p$ distribution is equivalent to adjusting the Hamiltonian \cite{goldstein1950classical}.
\begin{equation*}
    H_{c}(\mathbf{q}, \mathbf{p}) = H_{o}(\mathbf{q}, \mathbf{p}) - k_B T \sum_s \lambda_s \hat{O}_s(\mathbf{q}, \mathbf{p})
\end{equation*}

To efficiently optimize the parameters $\theta$ , we address the challenge of computing gradients through long-time Molecular Dynamics (MD) trajectories. Standard backpropagation-through-time is computationally infeasible here due to the exploding memory requirements of storing the computation graph for high-dimensional many-body systems. 

Therefore, we implement the Adjoint Sensitivity Method within the framework of Neural Ordinary Differential Equations (Neural ODEs) \cite{chen2020learning}. Instead of standard backpropagation, we solve the adjoint equation backward in time. We define an augmented state $[\mathbf{z}(t), \mathbf{a}(t)]$, where $\mathbf{y}(t)$ represents the phase space state (positions $\mathbf{q}$, momenta $\mathbf{p}$, and thermostat/barostat variables) and $\mathbf{a}(t) = \partial L / \partial \mathbf{y}(t)$ is the adjoint state. The optimization proceeds in two steps:
\begin{enumerate}
    \item \textbf{Forward Pass:} In the forward pass, the system is propagated over a finite simulation window in the isothermal--isobaric ensemble using the Nose--Hoover Chain NPT Verlet integrator. Starting from the initial NPT state, the molecular trajectory is generated by integrating the continuous-time dynamics defined by the neural potential model. Atomic configurations sampled from this trajectory are then used to evaluate the simulated pair distribution function (PDF), which is compared against the experimental PDF to define the training objective. In this way, the supervision is imposed not on a terminal microscopic state, but on a trajectory-derived structural observable that encodes the underlying potential energy surface.
    \item \textbf{Backward Pass:} To compute gradients efficiently, we employ the adjoint sensitivity method and integrate an augmented adjoint system backward in time over the same simulation window. The augmented system contains the original dynamical states $$\mathbf{y}(t) = [\mathbf{v}(t), \mathbf{q}(t), \mathbf{p}_v(t), \boldsymbol{\eta}(t), \boldsymbol{\zeta}(t), \mathbf{h}(t)]^T$$, the adjoint variables associated with those states $$\mathbf{a}(t) = \frac{\partial \mathcal{L}}{\partial \mathbf{y}(t)}$$ and $$\frac{d\mathbf{a}}{dt} = -\left( \frac{\partial \mathcal{F}(\mathbf{y}(t), t; \boldsymbol{\theta})}{\partial \mathbf{y}} \right)^T \mathbf{a}(t)$$, meanwhile, the accumulated sensitivities with respect to time and trainable parameters. During the backward pass, the PDF-based loss is propagated through the continuous-time dynamics via vector--Jacobian products, allowing gradients to be accumulated without explicitly storing the full computational graph of the forward trajectory,which is totally different from traditional methods like Backpropagation, as FIG\ref{fig:Adjoint ODE methods}shows. This procedure enables the experimental PDF information to drive the optimization of the trainable neural correction term stacked on top of the fixed prior potential, thereby refining the effective potential energy surface in a physically informed manner.
\end{enumerate}
\begin{widetext}
\begin{center}
\includegraphics[width=\textwidth]{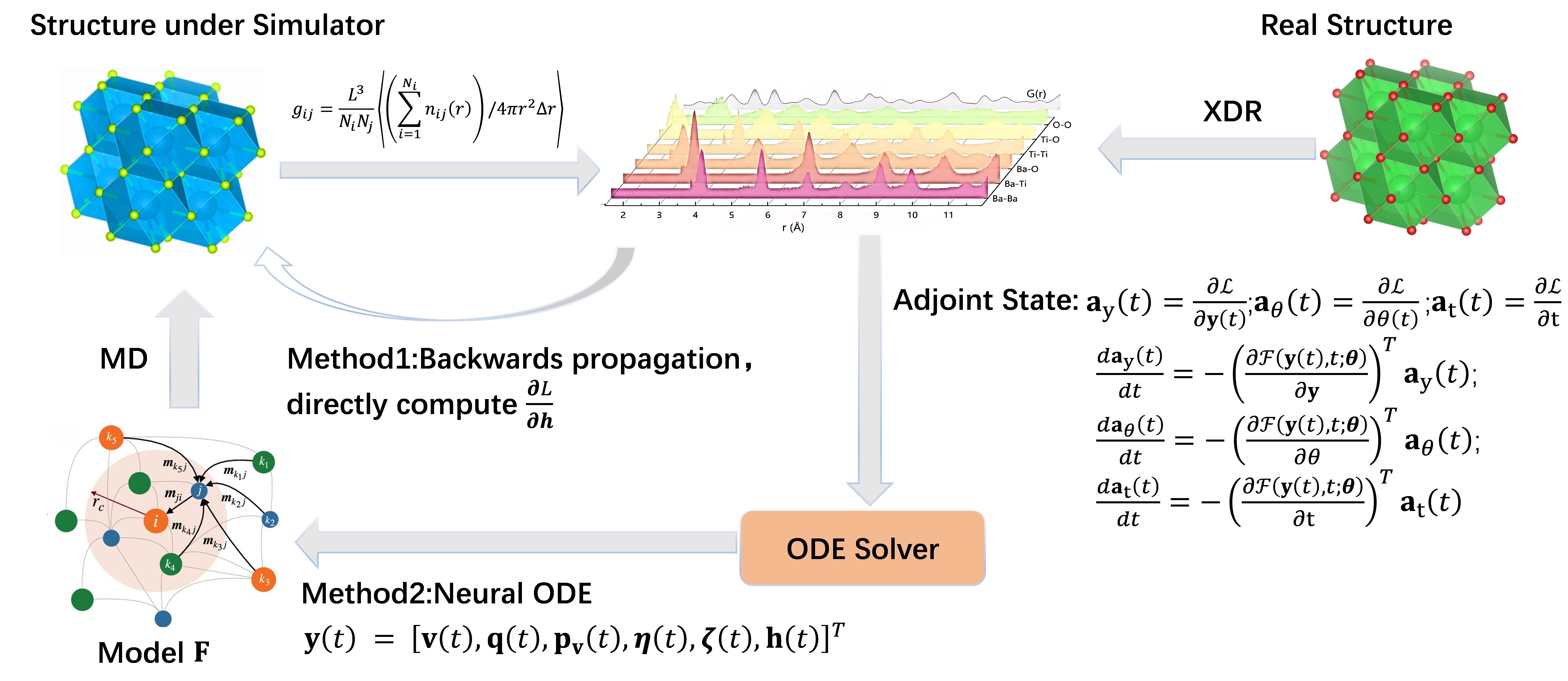}
\manualwidefigurecaption{fig:Adjoint ODE methods}{Presentation of Adjoint ODE methods.}
\end{center}
\end{widetext}

To couple adjoint dynamics with practical engineering optimization, we construct an augmented state that jointly carries four components: the original physical state $\mathbf{y}$, the state-adjoint variable $\mathbf{a}_{\mathbf{y}}$, the time-adjoint variable $a_t$, and the parameter-adjoint variable $\mathbf{a}_{\boldsymbol{\theta}}$. This design allows the solver to propagate physical trajectories and gradient information in one unified dynamical system.
\begin{equation}
\mathbf{Y}_{\mathrm{aug}}(t)
=
\begin{bmatrix}
\mathbf{y}(t)\\
\mathbf{a}_{\mathbf{y}}(t)\\
a_t(t)\\
\mathbf{a}_{\boldsymbol{\theta}}(t)
\end{bmatrix},
\quad
\mathbf{y}(t)=\begin{bmatrix}\mathbf{v}(t)\\\mathbf{q}(t)\\\mathbf{p}_v(t)\\\eta(t)\\\zeta(t)\\h(t)\end{bmatrix},
\quad
\mathbf{a}_{\mathbf{y}}(t)=\begin{bmatrix}\mathbf{a}_{\mathbf{v}}(t)\\\mathbf{a}_{\mathbf{q}}(t)\\\mathbf{a}_{\mathbf{p}_v}(t)\\a_{\eta}(t)\\a_{\zeta}(t)\\a_h(t)\end{bmatrix}.
\label{eq:aug_state_npt}
\end{equation}
\begin{equation}
\frac{d\mathbf{Y}_{\mathrm{aug}}(t)}{dt}
=
\begin{bmatrix}
\dot{\mathbf{y}}(t)\\
\dot{\mathbf{a}}_{\mathbf{y}}(t)\\
\dot{a}_t(t)\\
\dot{\mathbf{a}}_{\boldsymbol{\theta}}(t)
\end{bmatrix}
=
\begin{bmatrix}
\mathbf{g}_{\mathbf{y}}(t)\\
\mathbf{g}_{\mathbf{a}}(t)\\
g_t(t)\\
\mathbf{g}_{\boldsymbol{\theta}}(t)
\end{bmatrix},
\label{eq:aug_dynamics_npt}
\end{equation}
\begin{equation}
\begin{split}
\mathbf{g}_{\mathbf{y}}&=f\!\left(t,\mathbf{y},\boldsymbol{\theta}\right),\quad
\mathbf{g}_{\mathbf{a}}=-\mathbf{a}_{\mathbf{y}}^{\mathsf T}\frac{\partial f}{\partial \mathbf{y}},\\
g_t&=-\mathbf{a}_{\mathbf{y}}^{\mathsf T}\frac{\partial f}{\partial t},\quad
\mathbf{g}_{\boldsymbol{\theta}}=-\mathbf{a}_{\mathbf{y}}^{\mathsf T}\frac{\partial f}{\partial \boldsymbol{\theta}}.
\end{split}
\end{equation}
This approach allows for $O(1)$ memory cost with respect to integration time and ensures numerical stability when calculating gradients for high-dimensional tensor inputs (multi-particle systems) mapped to low-dimensional statistical outputs (PDF/RDF), effectively solving the complexity issues inherent in differentiable molecular dynamics.
 
\subsection{\label{subsec:dimenet_prior_ocvf}DimeNet as the Prior Model in OCVF}

In the present framework, DimeNet does not constitute the main methodological novelty by itself; rather, it provides the prior Hamiltonian $H_0$, namely the \emph{Model\_F} in Fig.~\ref{fig:Adjoint ODE methods}, upon which the observable-constrained correction is built. This choice is motivated by the work of Ouyang \emph{et al.}\cite{Ouyang2023}, who demonstrated that a refined and simplified DimeNet++-type message-passing neural network can serve as a high-fidelity energy model for perovskites and can be directly integrated into molecular dynamics simulations to reproduce the structural phase-transition sequence of BaTiO$_3$. Their study established both the architectural feasibility of DimeNet-based priors for ferroelectric perovskites and the importance of carefully calibrated prior parameters for obtaining realistic phase behavior.

Following this line, we treat the DimeNet-based model as a physically credible prior energy model rather than as the final target. In other words, the role of DimeNet here is to provide a transferable structural-energy backbone that already captures the basic transition skeleton of BaTiO$_3$, while the remaining discrepancy with experiment is corrected through OCVF. Specifically, the simulated PDF generated from the differentiable NPT dynamics is compared with the experimental PDF, and the resulting observational mismatch is backpropagated through the ODE solver to update the correction network parameters associated with Model\_F. In this way, the prior model is not discarded; instead, it is experimentally calibrated at the ensemble level.

A crucial point is that the learned object should not be interpreted as a bare Hamiltonian that explicitly changes with temperature. Rather, the DimeNet prior is supplemented by an experimentally constrained, temperature-dependent \emph{effective Hamiltonian family}, whose role is to reproduce the correct ensemble statistics and free-energy competition across temperature regimes. Thus, DimeNet supplies the prior framework, while OCVF and the PDF observer provide the variational bridge from the prior model to experimentally calibrated effective dynamics.
\subsection{\label{subsec:metric_function_ds}Metric Function ($D_s$)}
  The metric function $D_s$, used as the training loss, quantifies the discrepancy between the simulated observable $\mathfrak{O}_{sim,s}$ and the experimental observable $\mathfrak{O}_{exp,s}$. In statistical mechanics, this corresponds to a divergence between two probability distributions, $P$ (experimental) and $Q$ (simulated). A common choice is the Kullback--Leibler (KL) divergence (relative entropy) \cite{lin2002divergence}.
  However, for our application, the KL divergence $I(p_1 \| p_2)$ has two important drawbacks \cite{lin2002divergence}: (i) strict domain constraints---it is undefined when $p_2(x)=0$ while $p_1(x)\neq 0$ (i.e., $p_1$ must be absolutely continuous with respect to $p_2$); and (ii) unboundedness---there is no general upper bound in terms of the variational ($L_1$) distance, which can cause numerical instability. For PDF-based training, this issue is particularly severe: if the simulated PDF $g_{sim}$ assigns near-zero probability where the experimental PDF $g_{obs}$ is nonzero, the KL term can become arbitrarily large. To mitigate this pathology, we additionally monitor a symmetric log-based discrepancy, analogous to a Jensen--Shannon-type divergence, as an auxiliary bounded diagnostic quantity. For our PDF application, the JS loss monitor is:
  \begin{equation}
  L_{JS} = -\sum_{r} \left[ g_{obs}(r) \log\left(\frac{g_m(r)}{g_{obs}(r)}\right) + g_{sim}(r) \log\left(\frac{g_m(r)}{g_{sim}(r)}\right) \right]
  \end{equation}
  where $g_m = \frac{1}{2}(g_{obs} + g_{sim})$. This expression is, in fact, the sum of two KL divergences (termed $K$ divergence in the original paper ) :
  \begin{equation}
      L_{JS} = D_{KL}(g_{obs} || g_m) + D_{KL}(g_{sim} || g_m)
  \end{equation}
  The JS divergence (termed the $L$ divergence when $\pi_1=\pi_2=0.5$) has several desirable properties, including non-negativity, symmetry, finiteness, and boundedness, and is therefore used as an auxiliary monitor for numerical stability during training\cite{raj2008non}. For PDF-constrained optimization, however, the primary objective is defined as a physically weighted squared discrepancy between the simulated and experimental pair distribution functions, rather than the Jensen--Shannon divergence. Specifically, deviations $\Delta g(r)$ at larger distances $r$ are weighted more strongly to emphasize errors in long-range atomic correlations:
  \begin{widetext}
  \begin{equation}
  D = \sum_r \left[ 4\pi\rho_0 r^2 (g_{sim}(r) - g_{obs}(r))^2 \Delta r \right] \approx \int 4\pi\rho_0 r^2 (g_{sim}(r) - g_{obs}(r))^2 dr
  \end{equation}  
  \end{widetext}
\subsection{Differentiable Forward Model ($F_s$)}
The differentiable forward model ($F_s$) establishes a mapping from the internal physical parameters of the complex system to the external observational data. Specifically, in this paper, $F_s$ is a map from the corrected potential energy surface (PES), $H_c$, to the simulated observable $\mathfrak{O}_{sim, s}$.This mapping is realized by a forward model based on a Neural Network Potential (NNP) and a Neural Ordinary Differential Equation (Neural ODE) \cite{rackauckas2019diffeqflux}. The core idea of a Neural ODE is to embed a differential equation solver, which itself encodes structural physical priors, directly into a neural network as a differentiable layer.In our framework, $F_s$ takes the corrected Hamiltonian $H_c$ (i.e., $H_o + \Delta H_\theta$) as its input, which defines the dynamics of the system. It then uses this Hamiltonian to compute the forces and stresses required to integrate the equations of motion \cite{frenkel2023understanding}. This approach is an extension of modern "bottom-up" NNP-driven simulations, such as Deep Potential Molecular Dynamics (DeePMD), which have proven that NNP-driven MD can achieve quantum-mechanical accuracy at a fraction of the computational cost \cite{zhang2018deep}.The model $F_s$ operates under the external simulation conditions $\mathbf{Z}_s$ (e.g., NPT ensemble) by numerically integrating the atomic trajectories over time. Finally, it computes the simulated PDF using a differentiable histogram function. This entire simulation process ($F_s$) is differentiable, allowing gradients to be backpropagated from the observable error back to the Hamiltonian parameters $\theta$. This is made computationally efficient by using the Adjoint Sensitivity Method to calculate the gradient of the ODE solver, as detailed in the literature \cite{rackauckas2019diffeqflux}.The goal of $F_s$ is to produce a simulated observable $\mathfrak{O}_{sim, s}$ that matches the experimental constraint $\mathfrak{O}_{exp, s}$, such that $\langle \hat{O}_s \rangle_{\rho_c} = \mathfrak{O}_{exp, s}$. The microscopic operator $\hat{O}_s$ for the PDF, under the experimental NPT ensemble $\rho_c$, is:
\begin{equation}
\quad \langle \hat{O}s \rangle = \left\langle \frac{1}{N\rho_c} \sum_{i \neq j} \hat\delta(r - |\mathbf{R}i - \mathbf{R}j|) \right\rangle_{E{c}, T}
\end{equation}This process is validated by established NNP-MD methods, which confirm that simulated Pair Distribution Function (PDFs, i.e., PDFs) are "reproduced with high accuracy" compared to their target data.The experimental BTO data used as constraints $\mathfrak{O}_{exp, s}$ in this work is as Fig.\ref{Fig:RDF_origin} displays in Appendix:$\mathfrak{O}{exp,(N,P=1bar,T)} \in {\mathfrak{O}{exp, (N,P,T)}}$

\subsection{Observation Error ($\frac{\partial D_s}{\partial \mathfrak{O}_{sim, s}}$)}
$\frac{\partial D_s}{\partial \mathfrak{O}_{sim, s}}$ computes the error between the simulated and experimental PDF. $\mathfrak{O}_{sim, s}$ is the simulated observable: the simulated PDF ($g$) obtained by sampling and averaging trajectories under the NPT ensemble. The simulation results (at various training epochs) are as FIG.\ref{Fig:RDF_traning} follows:
\begin{widetext}
\begin{center}
\includegraphics[width=\textwidth]{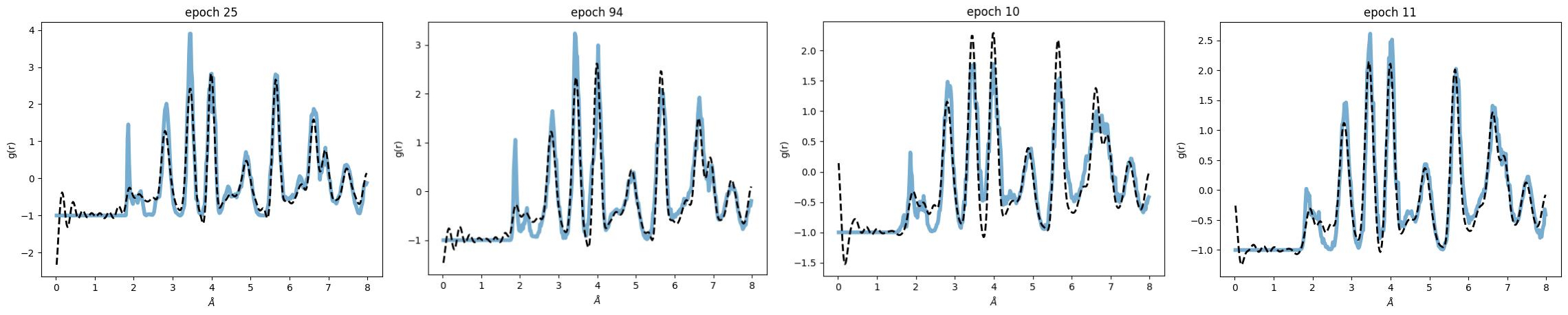}
\manualwidefigurecaption{Fig:RDF_traning}{Simulated PDFs: $\mathfrak{O}_{\mathrm{sim},(N,P=1\,\mathrm{bar},100\,\mathrm{K})}$, $\mathfrak{O}_{\mathrm{sim},(N,P=1\,\mathrm{bar},150\,\mathrm{K})}$, $\mathfrak{O}_{\mathrm{sim},(N,P=1\,\mathrm{bar},250\,\mathrm{K})}$, and $\mathfrak{O}_{\mathrm{sim},(N,P=1\,\mathrm{bar},300\,\mathrm{K})}$.}
\end{center}
\end{widetext}

In our specific model, the term $\frac{\partial F_s}{\partial H_{c}}$ represents the NPT integrator, which acts as the differentiable forward model $F_s$. It executes the ensemble simulation not in a vacuum, but under the strict NPT (isothermal-isobaric) ensemble conditions that mimic the experiment. This integrator takes the corrected Hamiltonian $H_{c}$ (i.e., $H_{o} + \Delta H_{\theta}$) as its input, which is called at every MD step to compute the forces and stresses driving the atomic motion. This entire process is a direct application of modern Machine Learning Potentials (MLPs) for atomistic simulations .The specific equations of motion for the NPT ensemble, which couple atomic coordinates $q$ and the cell matrix $h$ to thermostats and barostats (as implemented in NoseHooverChain\_NPT ), were rigorously derived by Martyna et al. to generate this exact ensemble \cite{martyna1994constant}. This integrator propagates the system's state variables ($q$ and $h$) using a reversible, time-reversible algorithm like NPT\_verlet\_update, which is a variation of the velocity Verlet algorithm adapted for the extended NPT phase space \cite{martyna1994constant, tuckerman1992reversible}.Solving for the gradient $\frac{\partial F_s}{\partial H_{c}}$ (the sensitivity of the final averaged observable to the potential) using classical forward methods is computationally intractable for a large number of parameters $\theta$ . Therefore, this framework innovatively uses the Adjoint Sensitivity Method (ASM). 

This approach is central to modern scientific machine learning for "backpropagating" through differential equation solvers, which can be viewed as Neural Ordinary Differential Equations (ODEs) \cite{rackauckas2019diffeqflux}.In this process, a new set of "adjoint" differential equations is derived, often via a Lagrangian formulation. These adjoint equations are then solved backward in time along the original MD trajectory . This method efficiently computes the complex Jacobian-vector product required for the gradient without needing to store the full computational graph, a process that is rigorously defined for Differential-Algebraic Equation (DAE) systems \cite{cao2003adjoint}.The physical meaning of $\frac{\partial F_s}{\partial H_{c}}$ is therefore the physical bridge connecting the statistical mechanical macroscopic observation error ($\delta g(r)_{NPT}$) with the microscopic potential energy surface (PES) model.Taking 300 K as an example, we clearly demonstrate via a three-dimensional image the correction of the Prior\_PES (potential energy surface trained by DFT) by PDF, as shown in FIG. \ref{Fig:Three dimension of correction under 300K}. From left to right, they are the potential energy correction of the Net at 300 K, the Prior PES at 300 K, and the ultimately formed potential energy surface.
Figure.\ref{Fig:correction of H} and Table \ref{tab:energy_corrections} in the Appendix together illustrate the two-dimensional PES correction and its corresponding numerical values in practice, showing how the differentiable forward model $F_s$ (driven by $\delta \mathfrak{O}{s}$ or $\delta g(r)$) corrects the prior potential energy surface $H_o$.
\begin{widetext}
\begin{center}
\includegraphics[width=\textwidth]{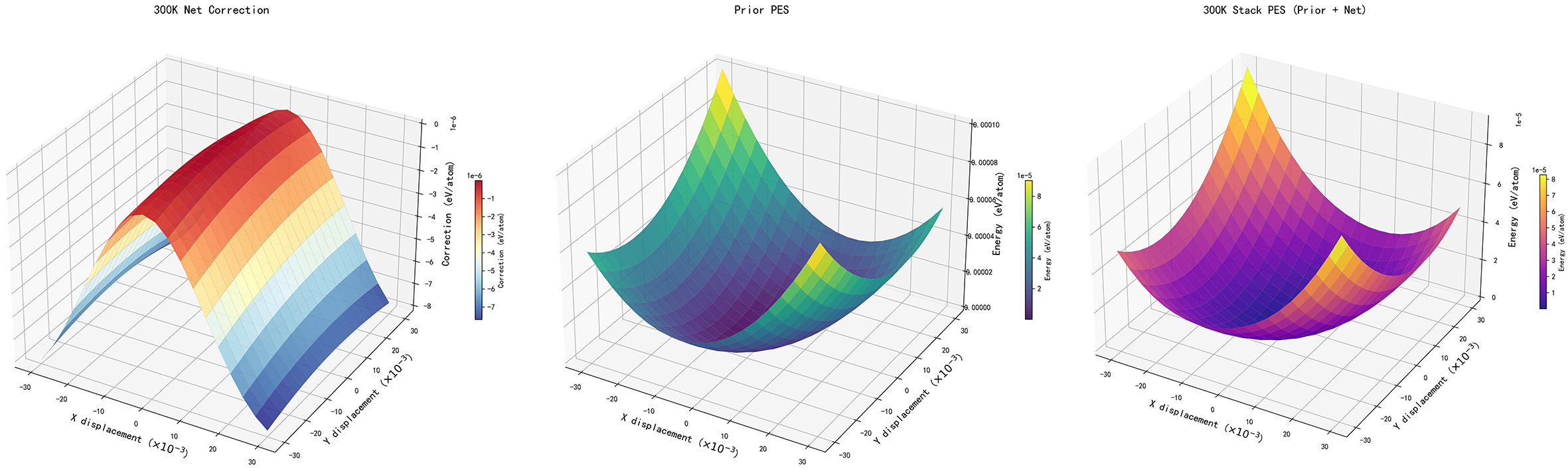}
\manualwidefigurecaption{Fig:Three dimension of correction under 300K}{Three-dimensional correction at 300 K.}
\end{center}
\end{widetext}

\subsection{From Deep Learning PES to Parameters $\lambda_{k}$}
The partial derivative of the energy correction term $\Delta H$ with respect to its own parameters $\lambda_{k}$ (represented as $\frac{\partial \Delta H_{\theta}}{\partial \theta}$) is automatically computed by the Automatic Differentiation (Autograd) engine of the deep learning framework.Functioning as a standard backpropagation process, its role is to coordinate with the optimizer: it determines exactly which weight parameters within the network need to be updated, and with what intensity. This is done to realize the specific energy variation $\Delta H$ required by the term $\frac{\partial F_s}{\partial H_{\text{c}}}$ (the sensitivity of the forward model $F_s$ with respect to the corrected Hamiltonian $H_c$).This process effectively forces the neural network to learn a "correction field" (the "flesh" $\Delta H_{\theta}$) on top of the "backbone" DFT potential ($H_o$). The optimizer updates the weights $\lambda_k$ not to match DFT energies, but to minimize the error between the simulated observables ($\langle \hat{O}_s \rangle$) and the experimental truths ($\mathfrak{D}_{exp,s}$).
\section{State-Aware Integration of Observable-Constrained Potentials via Mixture-of-Experts}

The fundamental challenge for the backbone potential $H_{o}$ (trained exclusively on 0 K DFT data) in predicting phase transitions lies in its purely bottom-up nature. While $H_{o}$ provides a highly accurate static potential energy surface and correctly describes local chemical environments, it struggles to capture the complex temperature-induced anharmonicities and collective structural shifts that drive macroscopic phase transitions in finite-temperature molecular dynamics (MD). Consequently, relying solely on the static $H_{o}$ leads to systematic inaccuracies in predicting critical phase transition temperatures (e.g., predicting O-R at 100 K, T-O at 275 K, and C-T at 475 K).

From a pragmatic algorithmic perspective, one can introduce a neural network correction term ($\Delta H_{\theta}^{s}$) trained specifically on experimental macroscopic observables (e.g., $PDF(T_i)$) at a specific thermodynamic state $s_i$. This observable-constrained correction exhibits exceptional local accuracy, as it explicitly forces the MD trajectory to sample the correct atomic configurations corresponding to that specific temperature. However, recent benchmark studies on machine learning interatomic potentials (MLIPs) caution against the naive deployment of such localized corrections. Extrapolating a single local Net far beyond its training thermodynamic distribution fundamentally fails to capture the global phase transition sequence and frequently leads to catastrophic unphysical MD trajectories, as the model explores phase spaces devoid of training data \cite{fu2022forces, raja2024stability}.

To capture the complete C-T-O-R phase transition sequence, it is technically intuitive to integrate multiple Expert Nets trained across a spectrum of temperatures. Nevertheless, a direct linear superposition of these local Nets onto the Prior ($H_{eff} = H_o + \sum \Delta H_{\theta}^{s_i}$) introduces non-physical force fields that severely destabilize the system (e.g., causing the BTO lattice to artificially collapse during simulation). Such brute-force addition fundamentally violates the extensivity of energy, artificially multiplying the interaction magnitudes in overlapping regions.
To explicitly demonstrate the necessity of our MoE integration strategy, we perform a comparative analysis evaluating a single-anchor correction against our dynamically stitched framework. As shown in Table \ref{tab:anchor_compare}, applying only a local correction trained at 300 K (Prior + Net(300 K)) partially improves the high-temperature Cubic-Tetragonal (C-T) transition, shifting the critical temperature from 475 K down to 425 K. However, this localized patch lacks thermodynamic transferability; it completely fails to correct the lower-temperature transitions, leaving the Orthorhombic-Rhombohedral (O-R) transition stranded at the Prior's erroneous 100 K. In stark contrast, the state-aware MoE model (Prior + $\sum_i G_i(s')\Delta H_{\theta_i}$) successfully propagates physical corrections across the entire continuous temperature spectrum, substantially aligning the full C-T-O-R sequence with experimental reality.

To resolve this and organically fuse the localized structural corrections onto the Prior backbone, we propose an Observable-Constrained Mixture-of-Experts (MoE) strategy \cite{zhang_balancenet}. We introduce a state-aware Gaussian gating mechanism $G(s, s')$ to dynamically assign weights to each local Expert Net based on the current simulation state $s'$. In this architecture, a local Net is maximally activated only when the instantaneous state $s'$ approaches its training anchor $s_i$—the regime where its structural constraints are most physically valid.

The resulting effective Hamiltonian driving the MD simulation is formulated as:
\begin{equation}
H_{eff}(q,p,s') = H_o(q,p) + \sum_{i=0}^{N} G(s_i, s') \Delta H_{\theta}^{s_i}(q,p)
\end{equation}

To ensure a continuous and smooth transition of the forces across the phase space and strictly conserve the physical energy scale, the gating coefficients $G(s_i, s')$ must adhere to the partition of unity ($\text{Max}_{i} G(s_i, s') = 1$). For a well-ordered sequence of discrete training conditions $s_0 < s_1 < \dots < s_N$, we define $G(s_i, s')$ as a normalized interpolation function based on Gaussian Radial Basis Functions (Gaussian RBF):
\begin{widetext}
 \begin{equation}
G(s_i, s') =\begin{cases}\delta(s'-s_0)(i=0), & s' < s_0 \quad (\text{lower bound extrapolation}) \\\delta(s'-s_N)(i=N), & s' > s_N \quad (\text{upper bound extrapolation}) \\\frac{w_i(s')}{w_i(s') + w_{i-1}(s')}, & s_{i-1} \le s' \le s_i \quad (\text{intermediate interpolation})
\end{cases}
\end{equation} 
\end{widetext}

where the weight term is $w_k(s') = \exp\left(-\frac{(s' - s_k)^2}{2\sigma^2}\right)$. The variance $\sigma$ strictly controls the broadening, preventing excessive overlap among adjacent Expert Nets. This analytical expression adopts the local approximation form of Nadaraya-Watson kernel regression, ensuring adaptive, smooth weighted mixing based on the thermodynamic "distance" between the current simulation state and the anchor states. This MoE design rigorously preserves the physical baseline of the Prior model ($H_o + 1 \times Net_{local}$) without inducing overlapping non-physical perturbations. While this gating mechanism can theoretically be anchored to variables like pressure $P$, our current framework anchors it to temperature $T$, corresponding to the temperature-driven structural evolutions in the NPT ensemble.

By employing this dynamically gated $H_{eff}$, the MD integrator smoothly navigates across multiple consecutive temperature intervals. As shown in FIG. \ref{fig:correction to violin plots—derived BTO Phase Transition} below, we have obtained significantly more accurate BTO phase transitions and structural parameters. Because overlaps and intrinsic fluctuations arise in macroscopic observables (lattice dimensions, angles, and kinetic energy) across continuous temperature sweeps, we construct violin plots based on the mean and variance of these observables. 

\begin{center}
\includegraphics[width=\linewidth]{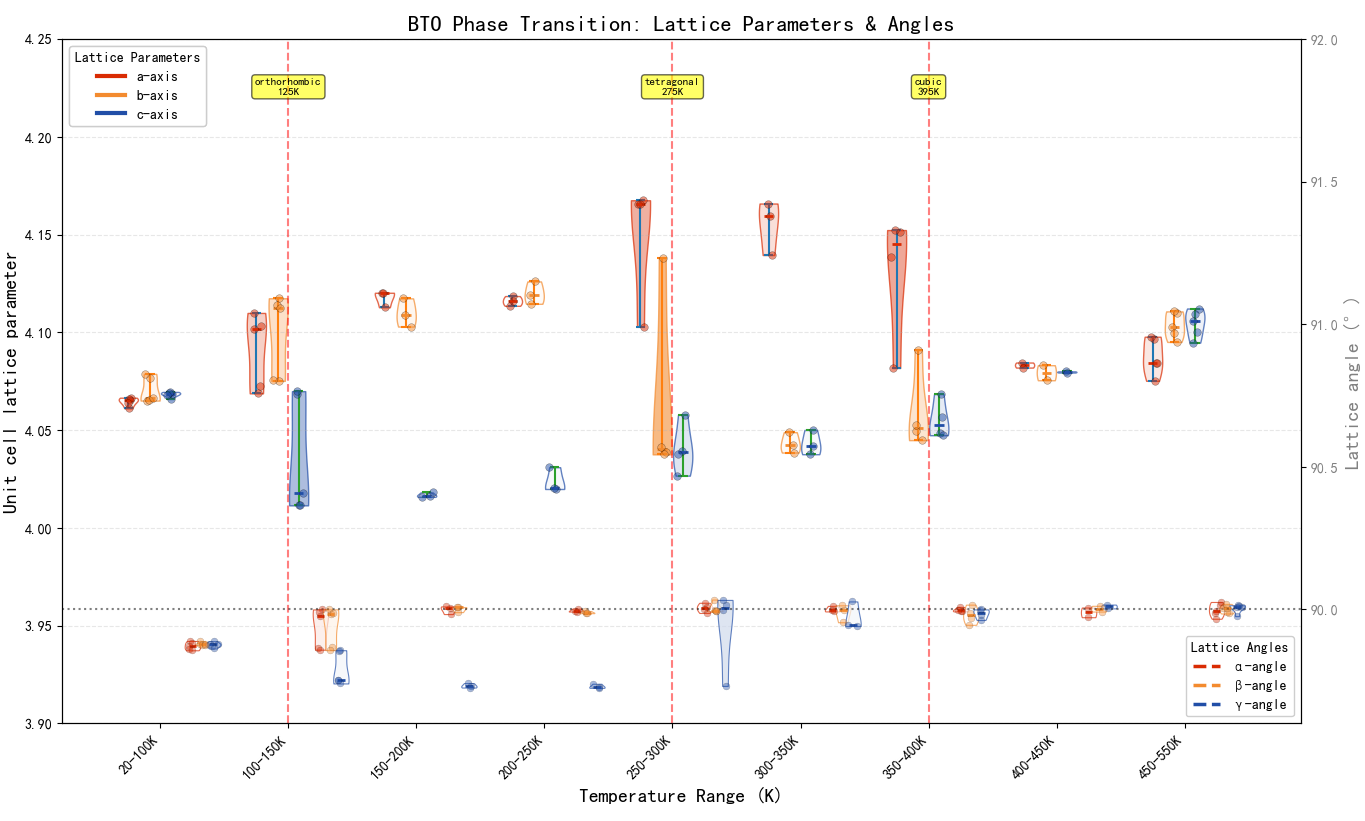}
\manualfigurecaption{fig:correction to violin plots—derived BTO Phase Transition}{Correction to BTO Phase Transition derived from violin plots of macroscopic observables.}
\end{center}

This visualization emphasizes our methodological contrast: while previous purely bottom-up studies often directly output singular lattice parameters, the proposed use of violin plots—derived from the statistical distributions of the MD trajectories—better captures uncertainties and state overlaps, perfectly aligning with the OCVF framework’s goal of infusing physical realism through macroscopic constraints. Through these violin plots, we can clearly observe that within non-phase-transition intervals, the variances of lattice parameters and angles are extremely small and stabilize around their respective means. 

\refstepcounter{figure}\label{fig:bto_lattice_transition_indicators}
\begin{center}
\includegraphics[width=0.85\linewidth]{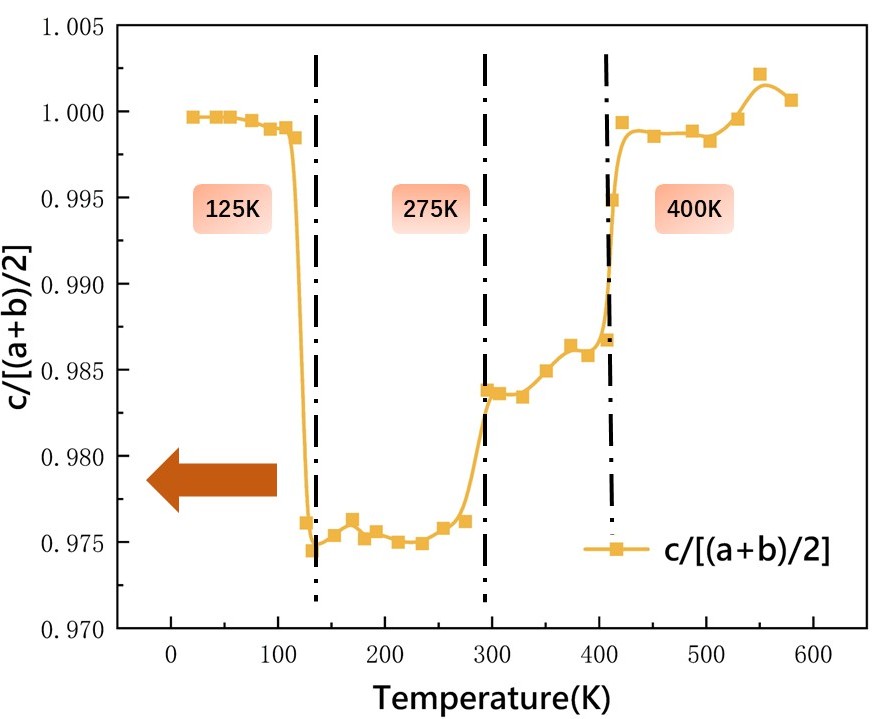}\\[6pt]
\includegraphics[width=0.85\linewidth]{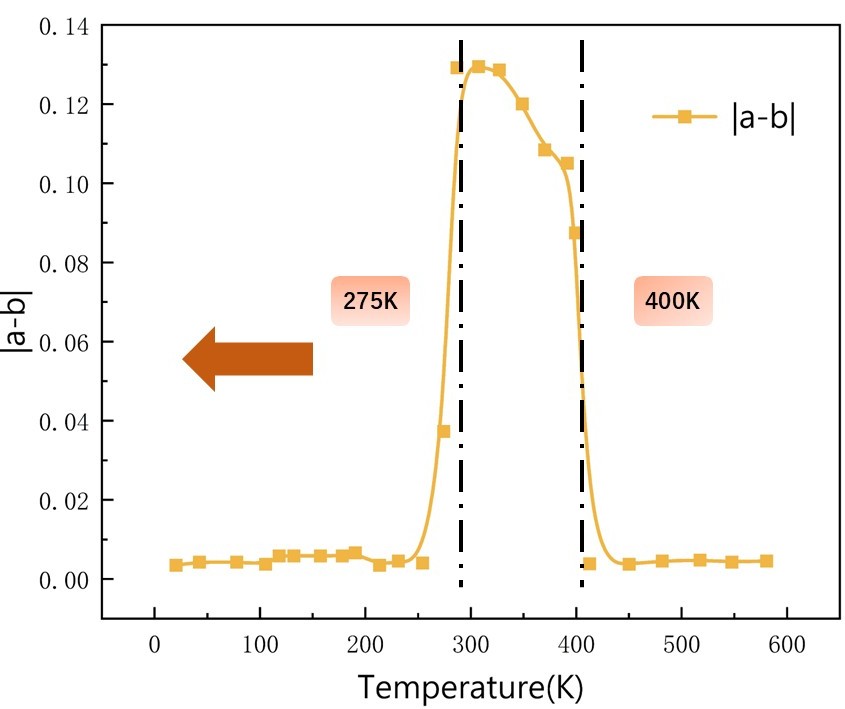}\par\smallskip
\small\textbf{FIG. \thefigure.} Lattice-transition indicators for BTO across different temperatures: the c-axis ratio (top) and the lattice-parameter difference $|a-b|$ (bottom).
\end{center}

In contrast, near the phase transition points, the variances of both lattice dimensions and angles exhibit an instantaneous increase of 400\%–550\%. As clearly demonstrated in FIG. \ref{fig:bto_lattice_transition_indicators}, the differences in lattice parameters and the ratios of their sums and differences distinctly reveal the phase transition temperatures as Cubic-Tetragonal (C-T) at 400 K, Tetragonal-Orthorhombic (T-O) at 275 K, and Orthorhombic-Rhombohedral (O-R) at 125 K.

Meanwhile, according to FIG. \ref{fig:BTO Polarization Transition}, which illustrates the $P_x, P_y$, and $P_z$ polarization transitions in [a] and the summation of total polarization in [b], together with Table \ref{tab:bto_stats}, which reports the specific numerical values of spontaneous polarization in the R, O, and T phases, the efficacy of the OCVF MoE integration is rigorously evaluated by comparing the phase transition temperatures and spontaneous polarization of $\mathrm{BaTiO_3}$ against the backbone ab initio model (Prior) and other standard density functional theory (DFT) functionals. As summarized in Table \ref{tab:phase_transition}, the Prior model, based on the PBE functional, exhibits well-documented systematic deviations: it significantly underestimates the critical temperatures for the rhombohedral-orthorhombic (R-O) and orthorhombic-tetragonal (O-T) transitions while overestimating the tetragonal-cubic (T-C) Curie temperature ($T_c$). Specifically, the Prior predicts a $T_c$ of 475 K, diverging from the experimental value of 403 K by nearly 20\%.

\begin{widetext}
\refstepcounter{figure}\label{fig:BTO Polarization Transition}
\begin{center}
\includegraphics[width=\textwidth]{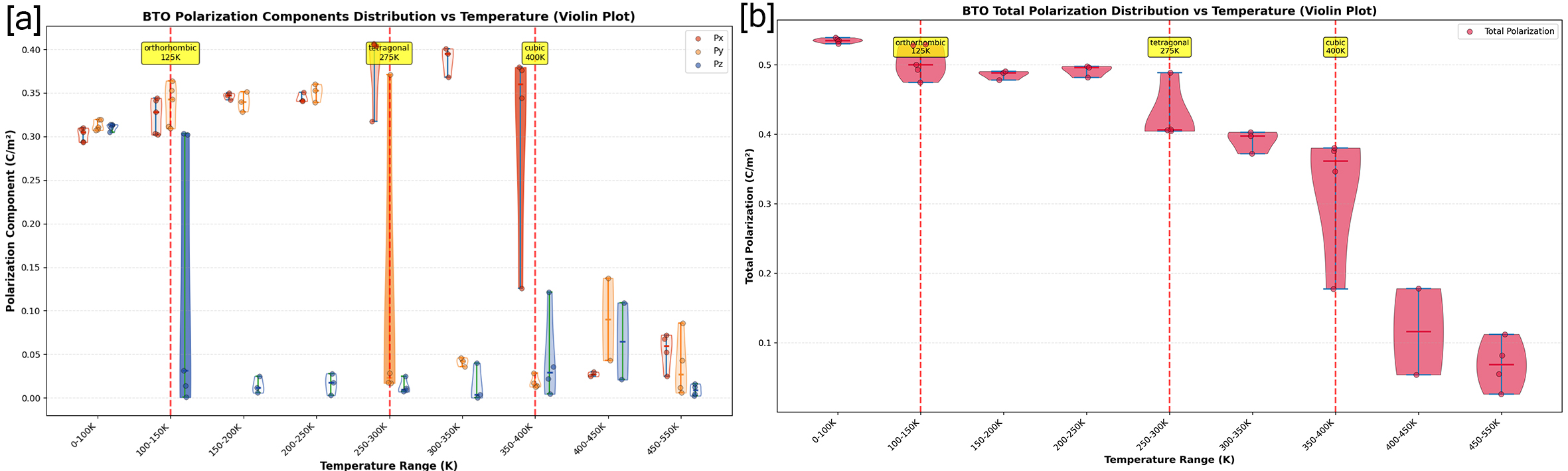}\par\smallskip
\small\textbf{FIG. \thefigure.} BTO polarization transition across different thermodynamic states.
\end{center}
\end{widetext}

The OCVF framework significantly enhances the prediction accuracy of macroscopic properties by organically fusing the state-specific corrections, thereby resolving the substantial deviations inherent in the pure Prior model. As Table \ref{tab:phase_transition} shows below, although the Prior model captures the C-T-O-R sequence qualitatively, it exhibits systematic offsets. For instance, its predicted Cubic-Tetragonal (C-T) transition at 475 K deviates significantly from the experimental 403 K. Furthermore, at low temperatures, the pure Prior model generates catastrophic non-physical results, indicating a lack of robustness during NPT simulations.

Improvements Achieved by OCVF:
\begin{itemize}
    \item \textbf{C-T Phase Transition (Cubic-Tetragonal):} The accuracy relative to the Prior model is improved by 95.8\%. OCVF predicts a transition temperature of approximately 400 K, closely aligning with the experimental value of 403 K.
    \item \textbf{O-R Phase Transition (Orthorhombic-Rhombohedral):} The accuracy is enhanced by 36.1\% compared to the Prior model. While the Prior model predicts 100 K, OCVF corrects this to 125 K (experimental value 183 K).
\end{itemize}

\begin{widetext}
\manualwidetablecaption{tab:phase_transition}{Comparison of phase transition temperatures ($T_c$) and spontaneous polarization ($P_s$) for different methods.}
\begin{center}
\begin{ruledtabular}
\begin{tabular*}{\textwidth}{l@{\extracolsep{\fill}}cccccc}
Method & \multicolumn{3}{c}{$T_c$ (K)} & \multicolumn{3}{c}{$P_s$ ($\mathrm{C/m^2}$)} \\
\cline{2-4} \cline{5-7}
 & O-R & T-O & C-T & R & O & T \\
\hline
OCVF & 125 & 275 & 400 & 0.52 & 0.46 & 0.36 \\
MPNN(PBE) & 100 & 275 & 475 & 0.54 & 0.5 & 0.41 \\
EH(LDA) & $200 \pm 10$ & $232 \pm 2$ & $296 \pm 1$ & 0.43 & 0.35 & 0.28 \\
EH(WC) & 102 & 160 & 288 &  &  &  \\
EH(SCAN) & 111 & 141 & 213 &  &  &  \\
GAP(PBEsol) & $18.6 \pm 0.4$ & $91.4 \pm 0.5$ & $182 \pm 0.7$ &  &  &  \\
Experiments & 183 & 278 & 403 & 0.33 & 0.36 & 0.27 \\
\end{tabular*}
\end{ruledtabular}
\end{center}
\end{widetext}
\manualtablecaption{tab:anchor_compare}{Comparison of phase-transition temperatures predicted by the prior model, a single-anchor correction, and the dynamically stitched MoE model.}
\begin{center}
\begin{ruledtabular}
\begin{tabular}{lccc}
Model & O--R (K) & T--O (K) & C--T (K) \\
\hline
Prior only & 100 & 275 & 475 \\
Prior + Net(300 K) & 100 & 275 & 425 \\
Prior+$\sum_i \Delta H_{\theta_i}$&Nan&Nan&Nan\\
Prior + $\sum_i G_i(s')\Delta H_{\theta_i}$ & 125 & 275 & 400 \\
Experiment & 183 & 278 & 403 \\
\end{tabular}
\end{ruledtabular}
\end{center}
OCVF successfully predicts the complete phase transition sequence accurately and eliminates the non-physical behaviors observed in the Prior model at low temperatures. In the rhombohedral phase, the prediction accuracy of the lattice structure is improved by 55.6\% compared to the Prior model. This enhancement demonstrates the framework's ability to calibrate microscopic interactions robustly via dynamically weighted macroscopic observables. 

The Prior model exhibits severe distortion when simulating polarization phase transitions under the NVT ensemble, as highlighted in FIG. 1. OCVF addresses this by dynamically correcting the forces in the Hamiltonian, ensuring the simulated structural landscape aligns with experimental reality. Although tabular data (e.g., Table \ref{tab:phase_transition}) indicate that the spontaneous polarization ($P_s$) predicted by OCVF (e.g., 0.52 $\mathrm{C/m^2}$ in the R phase) remains higher than the experimental value (0.33 $\mathrm{C/m^2}$), the physical consistency of the phase transition sequence is markedly improved, completely avoiding the non-physical trajectory oscillations inherent in the pure DFT model under expanded ensembles.

\appendix

\section{Derivation of the Constrained Ensemble}

To solve the variational problem defined in the main text, we use the method of Lagrange multipliers. We construct the Lagrangian functional as follows:
\begin{equation}
\begin{split}
\mathcal{L}[\rho] &= D_{\text{KL}} \\
&\quad - \sum_s \lambda_s \left( \int \rho_c \hat{O}_s \, d\mathbf{q} \, d\mathbf{p} - \mathfrak{O}_{\text{exp}, s} \right) \\
&\quad - \mu \left( \int \rho_c \, d\mathbf{q} \, d\mathbf{p} - 1 \right)
\end{split}
\end{equation}
Minimizing this functional with respect to $\rho_c$ requires setting the functional derivative to zero:
\begin{equation}
\frac{\delta \mathcal{L}}{\delta \rho_c} = 0
\end{equation}
This leads to the condition:
\begin{equation}
\log\rho + 1 - \log\rho_{o} - 1 - \sum_s \lambda_s \hat{O}s - \mu = 0
\end{equation}
Solving for $\rho{c}$, we get:
\begin{equation}
\rho_{c}(\mathbf{q}, \mathbf{p}) = \rho_{o}(\mathbf{q}, \mathbf{p}) \cdot \exp\left(\sum_s \lambda_s \hat{O}_s(\mathbf{q}, \mathbf{p})\right) \cdot e^{\mu}
\end{equation}
Substituting the definition of the prior distribution $\rho_{o} = \frac{1}{\mathcal{Z}_{o}} \exp(-\beta H_{o})$, and absorbing normalization constants into a new constant $\mathcal{Z}_{c}$, we arrive at the final form of the corrected ensemble:
\begin{equation}
\rho_{c}(\mathbf{q}, \mathbf{p}) = \frac{1}{\mathcal{Z}{c}} \exp\left( -\beta H{o} + \sum_s \lambda_s \hat{O}s \right)
\end{equation}
Comparing this to the standard Gibbs form $\rho \propto \exp(-\beta H{c})$, we identify the corrected Hamiltonian relation used in the main text: $-\beta H_{c} = -\beta H_{o} + \sum_s \lambda_s \hat{O}_s$.

\section{2-dimension of PES correction}
\begin{widetext}
\begin{center}
\includegraphics[width=\textwidth]{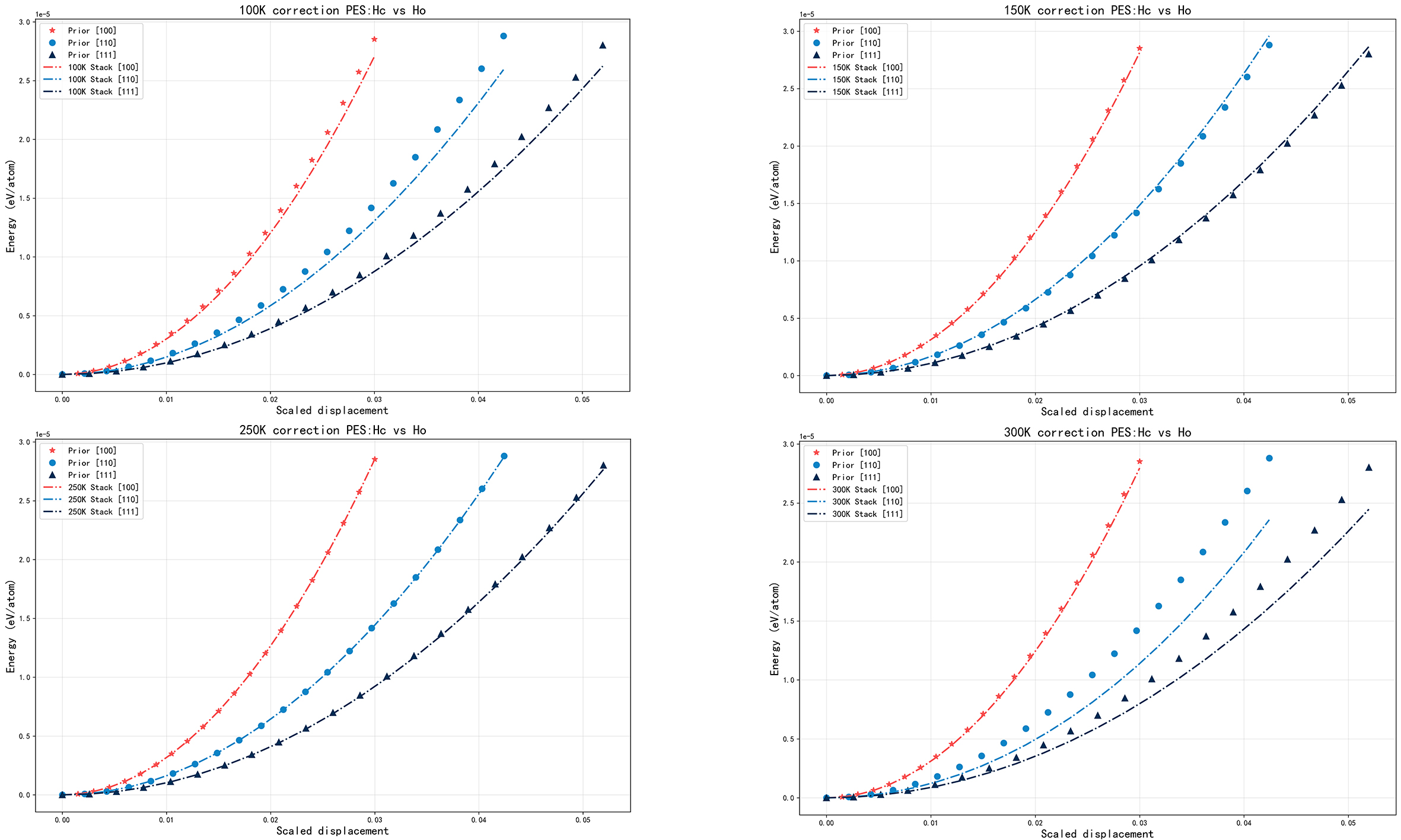}
\manualwidefigurecaption{Fig:correction of H}{Corrections to $H_o$ induced by $\partial F_{(N,P=1\,\mathrm{bar},T=100\,\mathrm{K})}/\partial H_{\mathrm{c}}$, $\partial F_{(N,P=1\,\mathrm{bar},T=150\,\mathrm{K})}/\partial H_{\mathrm{c}}$, $\partial F_{(N,P=1\,\mathrm{bar},T=250\,\mathrm{K})}/\partial H_{\mathrm{c}}$, and $\partial F_{(N,P=1\,\mathrm{bar},T=300\,\mathrm{K})}/\partial H_{\mathrm{c}}$.}
\end{center}
\end{widetext}
\begin{widetext}
\manualwidetablecaption{tab:energy_corrections}{Energy corrections (eV/atom) provided by the Net model relative to the Prior across different temperatures and crystallographic orientations. The table lists the maximum (Max) and root-mean-square (RMS) correction values.}
\begin{center}
\resizebox{\textwidth}{!}{%
\begin{tabular}{lcccccc}
  \toprule
  & \multicolumn{2}{c}{[100]} & \multicolumn{2}{c}{[110]} & \multicolumn{2}{c}{[111]} \\
  \cmidrule(lr){2-3} \cmidrule(lr){4-5} \cmidrule(lr){6-7}
  Temperature (K) & Max (eV/atom) & RMS (eV/atom) & Max (eV/atom) & RMS (eV/atom) & Max (eV/atom) & RMS (eV/atom) \\
  \midrule
  100 & $-1.47 \times 10^{-6}$ & $6.79 \times 10^{-7}$ & $-2.87 \times 10^{-6}$ & $1.32 \times 10^{-6}$ & $-1.79 \times 10^{-6}$ & $8.21 \times 10^{-7}$ \\
  150 & $-3.76 \times 10^{-7}$ & $1.75 \times 10^{-7}$ & 0 & $3.52 \times 10^{-7}$ & 0 & $2.86 \times 10^{-7}$ \\
  250 & 0 & $4.00 \times 10^{-9}$ & $-7.90 \times 10^{-8}$ & $3.87 \times 10^{-8}$ & $-3.68 \times 10^{-7}$ & $1.73 \times 10^{-7}$ \\
  300 & $-5.48 \times 10^{-7}$ & $2.56 \times 10^{-7}$ & $-5.23 \times 10^{-6}$ & $2.65 \times 10^{-6}$ & $-3.56 \times 10^{-6}$ & $1.74 \times 10^{-6}$ \\
  \bottomrule
\end{tabular}%
}
\end{center}
\end{widetext}
\section{PDF experimental data}
\begin{widetext}
\begin{center}
\includegraphics[width=\textwidth]{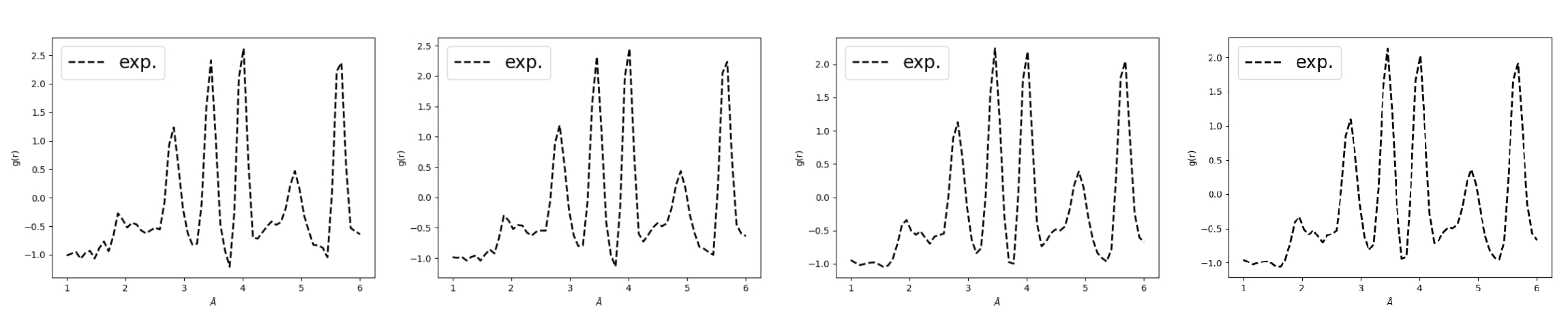}
\manualwidefigurecaption{Fig:RDF_origin}{Experimental PDF results: $\mathfrak{O}_{\mathrm{exp},(N,P=1\,\mathrm{bar},100\,\mathrm{K})}$, $\mathfrak{O}_{\mathrm{exp},(N,P=1\,\mathrm{bar},150\,\mathrm{K})}$, $\mathfrak{O}_{\mathrm{exp},(N,P=1\,\mathrm{bar},250\,\mathrm{K})}$, and $\mathfrak{O}_{\mathrm{exp},(N,P=1\,\mathrm{bar},300\,\mathrm{K})}$.}
\end{center}
\end{widetext}
\section{Results of BTO under OCVF correction}
\manualtablecaption{tab:bto_stats}{Statistical analysis of spontaneous polarization ($P_s$) for different phases of BTO. The units for mean and standard deviation are $\mathrm{C/m^2}$, and for variance is $(\mathrm{C/m^2})^2$.}
\begin{center}
\begin{ruledtabular}
\begin{tabular}{ccD{.}{.}{1.6}D{.}{.}{1.6}D{.}{.}{1.6}}
Phase & \multicolumn{1}{c}{Temp. Range (K)} & \multicolumn{1}{c}{Mean} & \multicolumn{1}{c}{Std. Dev.} & \multicolumn{1}{c}{Variance} \\
\hline
R & < 125    & 0.528940 & 0.011474 & 0.000132 \\
O & 125--300 & 0.467083 & 0.036032 & 0.001298 \\
T & 275--400 & 0.362768 & 0.068144 & 0.004644 \\
\end{tabular}
\end{ruledtabular}
\end{center}

\begin{acknowledgments}
This work is financially supported by “the Fundamental Research Funds for the Central Uni-versities”  of China (NSFC, Grant No. 11974268).
\end{acknowledgments}
\bibliography{aipsamp}
\end{document}